\documentstyle[twoside,fleqn,epsf,espcrc2]{article}

\title{Status of the {\it QCDSP} project}

\author{Dong~Chen,\address{Columbia University, New
York, NY}
\thanks{Research supported in part by the U.S. Dept.\
of Energy.}
\thanks{Current address: CTP-LNS, MIT, Cambridge, MA.}
Ping~Chen,$^{\rm a\;*}$
Norman~H.~Christ,$^{\rm a\;*}$
Robert~G.~Edwards,\address{SCRI, Florida State
University, Tallahassee, FL}$\;^*$
George~R.~Fleming,$^{\rm a\;*}$
Alan~Gara,\address{Nevis Laboratories, Columbia
University, Irvington, NY}
\thanks{Research supported in part by the National
Science Foundation.}
Sten~Hansen,\address{Fermi National Accelerator
Laboratory, Batavia, IL}
Chulwoo~Jung,$^{\rm a\;*}$
Adrian~L.~Kaehler,$^{\rm a\;*}$
Anthony~D.~Kennedy,$^{\rm b\;*}$
Gregory~W.~Kilcup,\address{Ohio State University,
Columbus, OH }$\;^{*}$ 
Yubing~Luo,$^{\rm a\;*}$
Catalin~I.~Malureanu,$^{\rm a\;*}$
Robert~D.~Mawhinney,$^{\rm a\;*}$
John~Parsons,$^{\rm c\;\ddagger}$
ChengZhong~Sui,$^{\rm a\;*}$
Pavlos~M.~Vranas$^{\rm a\;*}$\thanks{Current address:
Physics Dept., University of Illinois, Urbana, IL
61801}  and
Yuri~Zhestkov$^{\rm a\;*}$}

\begin{document}

\begin{abstract}
We describe the completed 8,192-node, 0.4Tflops machine
at Columbia as well as the 12,288-node, 0.6Tflops
machine assembled at the RIKEN Brookhaven Research
Center.  Present performance as well as our experience
in commissioning these large machines is presented.  We
outline our on-going physics program and explain how
the configuration of the machine is varied to support a
wide range of lattice QCD problems, requiring a variety
of machine sizes.  Finally a brief discussion is given
of future prospects for large-scale lattice QCD
machines.
\end{abstract}

\maketitle

\section{INTRODUCTION}
\label{sec:introduction}

The large computational requirements of lat\-tice QCD
coupled with the enormous cost/per\-formance advantages
that can be obtained with specially configured computer
hardware have encouraged the design and construction of
a variety of purpose-built machines over the past nearly
18 years.  The {\it QCDSP} machines now being completed
by our collaboration\cite{previous} represent a
continued development in this direction.

The most recent, 12,288-node machine at the RIKEN
Brookhaven Research Center has a construction cost of
approximately \$1.8M and a peak speed of 0.6Tflops or a
cost per peak performance of \$3/Mflops.  At present,
our most efficient routine inverts the Wilson Dirac
operator.  A complete program which carries out a
Wilson-fermion, hybrid Monte Carlo evolution achieves
about 25\% efficiency on a lattice volume of $4^4$
sites per node which corresponds to a dollar per
delivered Mflops figure of \$13.6/Mflops\cite{SC98}.

Achieving this level of performance required
considerable care when writing the routine which
applies the Wilson Dirac operator to a spinor field. 
(We expect that with more effort, this efficiency may
increase somewhat further and that our staggered code
will achieve a similar level of performance.)  This
high performance code is written in assembly language
and uses many of the special hardware features provided
to boost efficiency.  However, the bulk of the
conjugate gradient code which applies this Dirac
operator, as well as the hybrid Monte Carlo evolution
which uses the conjugate gradient inverter, is written
in $C^{++}$ and can be relatively easily understood and
modified.  Thus, this 25\% efficiency is achieved in a
``user friendly'',  $C^{++}$ physics software
environment in which the carefully coded, 
high-performance routines are easily called by the
high-level code---code which is no more difficult to
write than that in a more standard environment.

\section{PROJECT STATUS}
\label{sec:status}

\begin{figure}[t]
\vskip 0.1in
\epsfxsize=2.9in
\epsfbox{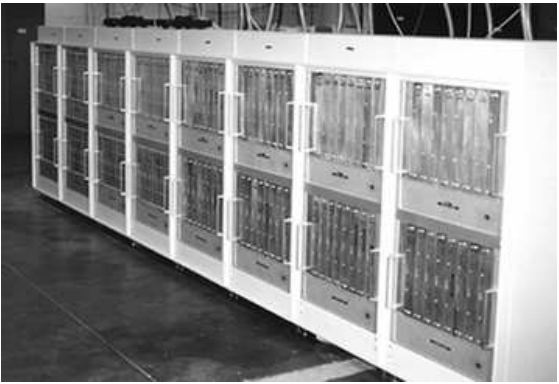}
\vskip -.2in
\caption{The 8,192-node, 0.4Tflops peak speed, {\it
QCDSP} machine running at Columbia since 4/98.}
\label{fig:128mb}
\vskip -0.3in
\end{figure}

We have now completed {\it QCDSP} machines at four of
the five sites listed in Table~\ref{tab:machines}. The
machine at the RIKEN Brookhaven Research Center is now
completely assembled, successfully runs our basic
diagnostic code and is in a final debugging mode.  The
machines at the other four locations have been running
production code for a number of months.  In particular,
the 8,192-node machine at Columbia
(Fig.~\ref{fig:128mb}) has been running quite reliably
in production mode since April 1998.  There is a
hardware failure roughly every two weeks which can
usually be repaired by simply replacing a faulty
daughter board, although on occasion more subtle errors
need to be diagnosed.  We anticipate that even this
error rate will decrease as a period of infant
mortality passes.

\begin{table}[t]
\setlength{\tabcolsep}{1.5pc}
\newlength{\digitwidth} \settowidth{\digitwidth}{\rm 0}
\catcode`?=\active \def?{\kern\digitwidth}
\caption{Locations of the five installed{\it QCDSP}
machines.}
\label{tab:machines}
\begin{tabular}{@{}lrr@{}}
\hline
Location       & Number  & Peak     \\
               & of Nodes& Speed    \\
\hline
RIKEN/BNL      & 12,288  & 0.6Tflops\\
Columbia       & 8,196   & 0.4Tflops\\
SCRI/FSU       & 1,024   & 50Gflops \\
OSU            & 128     & 6.4Gflops\\
Wuppertal      & 64      & 3.2Gflops\\
\hline
\end{tabular}
\end{table}

\section{CONSTRUCTION OF THE LARGE MACHINES}
\label{sec:construction}

During the past year, we have gone from operating up to
eight motherboards in an air cooled crate to much
larger machines composed of many cabinets for which
water cooling is supplied.  In order to allow the dense
stacking of the 8-motherboard crates within these
larger machines, we have arranged a water cooled, heat
absorber below the fan tray that blows cooling air
through each individual crate.

The introduction of water cooling has caused some
problems.  The room humidity must be controlled to avoid
condensation on the cold piping within the machine. 
With this internally-provided cooling, these large
computers become essentially decoupled from the cooling
and safety systems routinely provided in the larger
room housing the machine.  As a result, we have
installed temperature- and smoke-sensitive safety
systems in the large machines at both Columbia and
Brookhaven.

\section{PHYSICS PROGRAM}
\label{sec:configuration}

The earlier computers at Columbia were each configured
as a single large machine and used to tackle large,
demanding problems that typically had been previous
explored on smaller machines.  However, we are now able
to exploit the easily configurable character of the
{\it QCDSP} architecture, to adjust the hardware
arrangement to support a variety of physics goals
stretching from initial studies of dynamical 
domain-wall fermion thermodynamics on small, $8^3\times4$
lattices on 128-node, 6.4Gflops machines (we are
presently running $\approx 9$ such machines) to a
continuation of a large, $N_f=0$, 2 and 4 staggered
fermion calculation on 2 or 3 2048-node, 100Gflops
machines.  As our physics objectives mature, we expect
to assemble even larger units, ultimately running the
most demanding problems on a single machine composed of
perhaps 50-75\% of all available hardware.  Our 8,192
machine is presently running as 17 separate computers
on a wide range of lattices.

\section{FUTURE ARCHITECTURES}
\label{sec:future}

\begin{figure}[t]
\vskip -0.9in
\epsfxsize=3.3in
\epsfbox{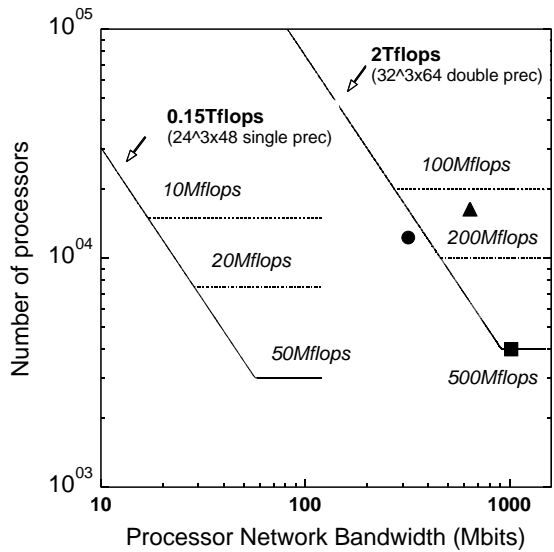}
\vskip -1.1in
\caption{The number of processors needed to sustain
2Tflops as a function of total network bandwidth
delivered to each node is shown by the curves on the
right for three different processors speeds.  The 
left-hand curves correspond to a machine which sustains
0.15Tflops.}
\label{fig:perf}
\vskip -0.3in
\end{figure}

Now we turn to a brief discussion of future machines.  
In order to keep the discussion reasonably concrete,
let us consider the design of a machine intended to
sustain 2.0Tflops on a $32^3 \times 64$ lattice.  We
assume that there will be demanding, full QCD
calculations that require a single large machine be
used to thermalize and evolve a single, relatively small
lattice.  Our performance estimates are based on 
those in \cite{QCDTeraflops}.  Two sets of curves are shown in
Fig.~\ref{fig:perf}.  The right-hand family shows the
number of processors required to sustain 2Tflops as a
function of network bandwidth on a $32^3\times 64$
lattice for preconditioned Wilson fermions and three
different processor speeds.  For comparison, the curves
on the left correspond to a 150Gflops machine with
slower individual processors.  (For simplicity the
effects of communication latency have been ignored in
this figure.)

For fixed speed processors and a fixed size problem, the number of
required processors falls as increasing bandwidth increases the
over-all efficiency of the machine.  Finally, when the communication
time falls below that required for the floating point portion of
calculation, the curve flattens and no longer depends on bandwidth.
The solid square corresponds to 4K, 0.5Gflops-sustained processors
with a 1Gbit network bandwidth available to each processor---a machine
that might be constructed from standard workstation and network
subsystems in a couple of years.  However, including modest $20\mu$sec
network latency will lower the sustained speed of this machine by
at least a factor of 2.  The solid triangle represents 16K,
0.15Gflops-sustained processors with 0.64Gbit network bandwidth, a
natural evolution of our {\it QCDSP} architecture with $2\times$
enhanced network.  Finally the solid circle shows the 0.15Tflops
sustained RIKEN/BNL machine.

This suggests that available technology
should permit either style of QCD-machine to be
constructed.  In deciding between these two approaches
one must compare the benefits of a standard
processor/workstation architecture to the significant
cost advantages achieved with a custom design.  It is
likely that a combination of the best of both
approaches will be possible.

\section{CONCLUSION}

This meeting marks the completion of the current set of
{\it QCDSP} machines under construction.  We now look
forward to a number of years during which these
significant resources will be used to increase our
understanding of both QCD and quantum field theory.

\end{document}